\documentclass[aps,prb,groupedaddress,showpacs,nobalancelastpage,onecolumn]{revtex4}
\usepackage{amssymb}
\usepackage{amsmath}
\usepackage{bm}
\usepackage{graphicx}
\usepackage{color}

\begin{document}

\title{FFLO state in thin superconducting films}
\author{A. Buzdin$^{1}$, Y. Matsuda$^{2,3}$, T. Shibauchi$^{2}$ }
\affiliation{$^{1}$ Institut Universitaire de France and Universit\'{e} Bordeaux I,
CPMOH, UMR 5798, 33405 Talence, France\\
$^{2}$ Department of Physics, Kyoto University, Sakyo-ku, Kyoto 606-8502,
Japan\\
$^{3}$ Institute for Solid State Physics, University of Tokyo, Kashiwa,
Chiba 277-8581, Japan}
\date{\today}

\begin{abstract}
We present the analysis of the inhomogeneous
Fulde-Ferrell-Larkin-Ovchinnikov (FFLO) superconducting state in thin
superconducting films in the parallel magnetic field. For the tetragonal
crystal symmetry (relevant to CeCoIn$_{5}$ - the most probable candidate for
the FFLO state formation) we predict a very peculiar in-plane angular
dependence of the FFLO critical field due to the orbital effect. In the
uniform superconducting state the critical field should be isotropic. The
magnetic field pins also the direction of the FFLO modulation\ permitting
thus to study the critical current anisotropy. Our calculations reveal a
strong critical current anisotropy in the FFLO state in sharp contrast with
the usual superconducting state. The predicted characteristic anisotropies
of the critical field and critical current may provide an unambiguous probe
of the FFLO phase formation.
\end{abstract}

\maketitle

\section{Introduction}

Recently the strong experimental evidence has been obtained in favor of the
existence of the inhomogeneous superconducting state in the heavy fermion
compound CeCoIn$_{5}$ (see ref.~\onlinecite{Matsuda} and references cited
therein). This state has been predicted a long time ago by Larkin and
Ovchinnikov \cite{LarkinOvchin}, and Fulde and Ferrell \cite{FuldeFerrel}
who demonstrated that in a pure ferromagnetic superconductor at low
temperature the superconductivity may be non-uniform (for a review see
refs.~ \onlinecite{Matsuda} and \onlinecite{Casalbuoni} ). The conditions of
such FFLO state formation are rather stringent. In CeCoIn$_{5}$ the FFLO
state exists owing to the paramagnetic pair-breaking effect which dominates
the orbital one. Moreover the superconductivity of CeCoIn$_{5}$ corresponds
to the clean limit. Note also that the magnetic-field-induced
superconductivity has been observed in the quasi two-dimensional organic
conductor $\lambda $-(BETS)$_{2}$FeCl$_{4}$, which may be another good
candidate for the FFLO state realization \cite{Uji}.

Although the upper critical field in CeCoIn$_{5\text{ }}$is dominated by the
paramagnetic effect, the orbital effect also plays an important role in this
compound leading to the vortex state formation. According to the theoretical
works \cite{GunterGrun,Bul1973,BuzBrison,BuzBrison2D,ShimaharaRainer} in the
FFLO state the vortex lattices may be very special and this circumstance
complicates an unambiguous identification of the FFLO phase. The ideal
system to study FFLO state without masking effect of the orbital field may
be a two dimensional superconductor in a parallel magnetic field. It is a
challenging experimental task to fabricate thin films of CeCoIn$_{5}$ with
thickness smaller than superconducting coherence length. There are
strong experimental evidences that CeCoIn$_{5}$\ is a $d$-wave
superconductor \cite{Matsuda}. The potential scattering is harmful for both
\ $d$-wave superconductivity \cite{Mineev} and FFLO phase \cite%
{Casalbuoni}. Therefore the films must have an atomically smooth surface and
be in the clean limit. However due to the recent progress of the multilayer
fabrication technology (in particular due to the epitaxial technique) such
systems would be created in a near future. In the present work we study
theoretically the properties of the FFLO\ state in superconducting film with
the thickness $d$ smaller than the superconducting coherence length $\xi $.
To be more specific, we assume the CeCoIn$_{5}$ multilayered structure with
the tetragonal crystal symmetry and $\mathbf{c}$ axis perpendicular to the
superconducting layers. Our approach is based on the modified
Ginzburg-Landau (MGL) theory \cite{BuzKach} which adequately describes the
FFLO state with a long wave length modulation - near the tricritical point
(TCP). This TCP is the meeting point of three transition lines separating
the normal, the uniform superconducting and the FFLO states. However
qualitatively our results may be applied to the whole region of the FFLO
state existence. The orbital effect for the magnetic field parallel to the
film is weak, and comparing with the bulk superconductor its contribution to
the upper critical field is reduced by the factor $\left( d/\xi _{c}\right)
^{2}\ll 1$, where $\xi _{c}$ is the superconducting coherence length along $%
\mathbf{c}$ axis. Then we may expect a much larger FFLO region on the $(H,T)$
phase diagram. In the present work, we demonstrate that small but finite
orbital effect leads to a characteristic in-plane anisotropy of the upper
critical field which may serve as a clear indication of the FFLO state
formation. Also the magnetic field lifts the degeneracy of the direction of
the FFLO modulation and permits to create a monodomain FFLO state. Moreover
our analysis shows that we may expect a pronounced anisotropy of the
superconducting critical current in FFLO state.

\section{Generalized Ginzburg-Landau functional}

The long period FFLO modulation near the superconducting transition can be
described by the MGL functional \cite{BuzKach} which in addition to the
usual gradient terms contains the higher derivatives of the superconducting
order parameter $\Psi $. The necessity to add the higher derivatives is
related to a special behavior of the coefficient on the usual gradient term
which goes through zero and becomes negative in the region of FFLO phase.
This circumstance makes the MGL theory qualitatively different from the
standard Ginzburg-Landau approach and is responsible for the peculiar
properties of the FFLO state. As an example, we note that in the
Ginzburg-Landau functional the higher derivatives describe the weak
non-local effects (see, for example, ref.~\onlinecite{Kogan}) which are
related to the details of the Fermi surface. In the FFLO phase these
non-local terms are of the primary importance and then making the properties
of the FFLO phase ultimately dependent on the details of the electronic
spectrum.

The MGL functional provides an adequate description of the FFLO state near
the trictitical point where the wave-vector of the FFLO modulation is small
but obtained results could be extrapolated qualitatively on the whole region
of the FFLO phase. In the case of pure paramagnetic effect this TCP
corresponds to $T^{\ast }=0.56T_{c0},$ $H^{\ast }=H(T^{\ast })=0.61\Delta
_{0}/\mu _{B}=1.05T_{c0}/\mu _{B}$, where $T_{c0}$ is the critical
temperature in the absence of the paramagnetic effect. The orbital effect
decreases $T^{\ast }$ and in CeCoIn$_{5}$ $T^{\ast }$ seems to be ($%
0.2-0.3)T_{c0}$ (ref.~\onlinecite{Matsuda}) depending on the field
orientation. In the bulk superconductors, the relative contribution of the
paramagnetic and orbital effects may be characterized by the Maki parameter $%
\alpha =\sqrt{2}\frac{H_{c2}^{orb}}{H_{p}^{{}}}$, where $H_{c2}^{orb}$ is
the orbital critical field extrapolated to $T=0$ from the slope of $%
H_{c2}(T) $ near $T_{c}$ and $H_{p}$ is a paramagnetic limit at $T=0.$ In
CeCoIn$_{5}$ the Maki parameter is large $\alpha \approx 5$ ensuring the
condition for the FFLO phase formation. The particularity of CeCoIn$_{5}$ is
that the superconducting transition is slightly first order \cite{Izawa}
below $\sim 0.4T_{c0}$ for the field $\mathbf{H\parallel ab}$ and below $%
\sim 0.3T_{c0}$ for $\mathbf{H\perp ab}$, and it remains first order at FFLO
transition \cite{Matsuda}. This intriguing behavior may be related with
rather pronounced magnetic fluctuations in this compound which is expected
to be at the vicinity of quantum critical point \cite{Bianchi}. More
generally the internal field $h$ acting on the electron spins have the
contribution coming from the Ce band, in addition to Zeeman's term $%
\widetilde{\mu }H$ (where $\widetilde{\mu }$ is effective electron magnetic
moment). If the exchange integral describing the interaction between this
band and the superconducting electrons is $I$, the contribution from the
polarized Ce atoms being $SI$, where $S$ is their relative magnetization.
Owing to the interband interaction, it would be a contribution to the Ce
polarizability from the electron susceptibility $\chi _{_{e}}$, which
changes at the superconducting transition $\chi _{_{e}}\sim \chi
_{_{e}}^{0}\left( 1-c^{/}\frac{\left\vert \Psi \right\vert ^{2}}{T_{c}^{2}}%
\right) ,$ where the constant $c^{/}\sim 1$ \cite{Bul-review}. In the result
the internal exchange field $h$ will acquire a correction $\sim
-IH\left\vert \Psi \right\vert ^{2}.$ If the exchange integral is positive
it will give a negative $\sim \left\vert \Psi \right\vert ^{4}$ contribution
in the Ginzburg-Landau functional. Note that for the negative exchange
integral we may have an inverse situation and then the superconducting
transition could be the second order one at all temperature. Therefore
depending on the sign of the exchange integral, the Ce band could favor the
first or second order superconducting transition.

Taking CeCoIn$_{5}$ in mind, we assume here that the superconducting
transition is weakly first order and then the MGL functional reads: 
\begin{eqnarray}
F_{G} &=&a(H,T)\left\vert \Psi \right\vert ^{2}-\alpha \left( \left\vert \Pi
_{x}\Psi \right\vert ^{2}+\left\vert \Pi _{y}\Psi \right\vert ^{2}\right)
+\beta \left( \left\vert \Pi _{x}^{2}\Psi \right\vert ^{2}+\left\vert \Pi
_{y}^{2}\Psi \right\vert ^{2}+\left\vert \Pi _{y}\Pi _{x}\Psi \right\vert
^{2}+\left\vert \Pi _{x}\Pi _{y}\Psi \right\vert ^{2}\right)  \label{MGL} \\
&&+\varepsilon \left( \Pi _{x}^{2}\Psi \left( \Pi _{y}^{2}\Psi \right)
^{\ast }+c.c.\right) -\frac{2b}{3}\left\vert \Psi \right\vert ^{4}\ +\frac{%
8\lambda }{15}\left\vert \Psi \right\vert ^{6},  \notag
\end{eqnarray}%
where $\Pi _{x}=\frac{\partial }{\partial x}-i2eA_{x}$ and $\Pi _{y}=\frac{%
\partial }{\partial y}-i2eA_{y}{}_{\text{ }}$, the superconducting film in $%
xy$ plane and $z$ axis perpendicular to the film. In the FFLO region the
coefficients $\alpha ,\beta >0$ and the choice $b,\lambda >0$ ensures the
first order transition. Magnetic field is directed along the film and makes
the angle $\theta $ with $x$ axis and then%
\begin{eqnarray*}
H_{y} &=&H\sin \theta ,\ H_{x}=H\cos \theta , \\
A_{y} &=&-zH\cos \theta ,\text{ }A_{x}=zH\sin \theta .
\end{eqnarray*}

The coefficient $a(H,T)$ vanishes at the line of the second order transition
to the uniform superconducting state and at fixed $H$ it may be written as $%
a=a_{0}(T-T_{cu}(H))$, where $T_{cu}(H)$ is the second order transition
temperature into the uniform state. Due to the small thickness of the film $%
d\ll \xi _{c}$, the superconducting order parameter is constant over its
thickness and then in (\ref{MGL}) we have omitted the derivatives on $z$.
The higher derivatives terms with the coefficient $\varepsilon $ describe
explicitly the difference between isotropic $s$-wave pairing model and the
real situation realized in the tetragonal crystals and/or with $d$-wave
pairing. In isotropic s-wave superconductor there is a degeneracy over
the direction of the FFLO modulation. For d-wave superconductor this
degeneracy is lifted and in 3D case the modulation vector is always directed
along the order parameter nodes \cite{Samokhin}. In 2D d-wave superconductor
at low temperature $T\lesssim 0.06T_{c}$\ a first order
re-orientational transition occurs to the state with a modulation vector
along the order parameter lobes \cite{Yang}. Note that in general $%
\varepsilon \sim \beta $ and the effect of anisotropy for the FFLO state
cannot be expected to be small. This is an important difference with the
standard Ginzburg-Landau theory where only the first rank tensors enter as a
gradient terms and then the cubic crystal structure (or tetragonal in $ab$
plane) is equivalent to the isotropic one. Then the form of the
Fermi-surface and the type of the superconducting pairing are both
equally important in determining the wave-vectors of FFLO modulation. 
For the FFLO transition the interplay between the Fermi surface
structure and the type of the superconductivity has been studied in refs.~ %
\onlinecite{Shimahara1} and \onlinecite{Shimahara2} \ on the basis of the
tight binding model and a very rich variety of the scenarios of the FFLO
transition was revealed. In general it may be demonstrated \cite{Brison}
that the effective mass approximation can be reduced to the isotropic model
by scaling transformation. However there are namely the deviations from the
elliptical Fermi-surface which are crucial to the adequate FFLO description.
Note that the tensor coefficients on the second derivatives terms in MGL are
given by the expression \cite{Houzet} \ $\beta _{ij}\sim \langle
v_{i}^{2}v_{j}^{2}\left\vert \psi (\mathbf{k})\right\vert ^{2}\rangle $,
where $v_{i}$ are the components of the Fermi velocity, $\psi (\mathbf{k})$
is the gap function and the averaging is performed over the Fermi surface.

\section{Orientational effect of the in-plane field. In-plane anisotropy of
the critical field.}

Let us consider first the quadratic terms in (\ref{MGL}) which depend on the
orientation of the FFLO modulation. If the transition is of the second order
(or weakly first order) the solution for the order parameter is of the form $%
\Psi (\mathbf{r})=f\cos (\mathbf{qr}).$

Without orbital effect 
\begin{equation}
\delta F\left( \varphi \right) \sim \left[ -\alpha q^{2}+\beta q^{4}+\frac{%
\varepsilon }{2}q^{4}\sin ^{2}2\varphi \right] \left\vert \Psi
_{q}\right\vert ^{2},  \label{F with no field}
\end{equation}%
where $\varphi $ is the angle between the FFLO modulation vector and \textbf{%
x}-axis. For $\varepsilon >0$ the minimum energy (and maximum critical
temperature) corresponds to $\varphi =0,$ $\pm \pi /2,\pi $ directions, i.e.
along the $x,y$ axis and the wave-vector $q_{0}=\sqrt{\frac{\alpha }{2\beta }%
}$. For $\varepsilon <0$ the minimum energy corresponds to $\varphi =\pm
3\pi /4,$ $\pm \pi /4$ directions, i.e. along the diagonals.

In the case of a thin superconducting film in a parallel field it is easy to
take into account the orbital effect - it is simply needed to average the
functional (\ref{MGL}) over the film thickness. The angular dependent part
is 
\begin{eqnarray}
\delta F\left( \varphi ,\theta \right) &\sim &\left\vert \Psi
_{q}\right\vert ^{2}\left\{ -\alpha \left( q^{2}+(2eH)^{2}\frac{d^{2}}{12}%
\right) +\beta \left( q^{4}+(2eH)^{4}\frac{d^{4}}{80}\right) +\frac{%
\varepsilon }{2}\left( q^{4}\sin ^{2}2\varphi +(2eH)^{4}\frac{d^{4}}{80}\sin
^{2}2\theta \right) +\right.  \label{AngDepPart} \\
&&\left. +2q^{2}(2eH)^{2}\frac{d^{2}}{12}\left[ 2\beta +\frac{\varepsilon }{2%
}-\left( \beta +\frac{\varepsilon }{4}\right) \cos (2\theta -2\varphi )+%
\frac{3\varepsilon }{4}\cos (2\theta +2\varphi )\right] \right\} .  \notag
\end{eqnarray}

Let us consider first the isotropic case $\varepsilon =0.$ Naturally the
properties of the superconducting system are not depending on the field
orientation itself and the angular dependent part has the form $-\beta
\left\vert \Psi _{q}\right\vert ^{2}q^{2}d^{2}(2eH)^{2}\cos (2\theta
-2\varphi )$ and the minimum of the energy is attended at $\theta =\varphi $%
, i.e. when the FFLO modulation is directed along the magnetic field. So the
magnetic field provides an orientational effect on the FFLO phase. In the
absence of other sources of anisotropy the resulting field dependent
contribution to the energy is isotropic in $xy$ plane. The quadratic over $H$
contribution vanishes for the wave-vector of the FFLO modulation $q=q_{0}=%
\sqrt{\frac{\alpha }{2\beta }}$. Then there is no linear diamagnetic
response in the FFLO phase $\delta F=\beta \left\vert \Psi _{q}\right\vert
^{2}q^{2}\frac{d^{2}}{6}(2eH)^{2}-\alpha (2eH)^{2}\frac{d^{2}}{12}\left\vert
\Psi _{q}\right\vert ^{2}=0$ . The resulting orbital field contribution $%
\sim \left\vert \Psi _{q}\right\vert ^{2}\beta (2eH)^{4}\frac{d^{4}}{180}$
is quartic over $H$ and then the diamagnetic moment is pretty small and
proportional to $H^{3}$.

In the more realistic case the anisotropy (crystalline and/or Cooper
pairing) plays a very important role and pins the orientation of the FFLO
modulation. Let us suppose for example that $\varepsilon >0$ and then in the
absence of the orbital effect the FFLO modulation vectors are along the $x,y$
axis. Note that in the case $\varepsilon <0$\ the rotation of
the $xy$\ axis by 45${{}^\circ}$
\ provides us the same functional (\ref{MGL}) with renormalized
coefficients $\widetilde{\varepsilon }$\ and $\widetilde{\beta }$%
\ but with $\widetilde{\varepsilon }>0.$ Therefore our analysis
presented below may be directly adapted to this case. If $\beta >\frac{%
\varepsilon }{2}$ and the angle of the magnetic field $\left\vert \theta
\right\vert <\pi /4$ then the direction of the wave vector $q$ will be close
to the $x$ axis while for $3\pi /4>\left\vert \theta \right\vert >\pi /4$
the modulation will be along $y$ axis. For $\beta <\frac{\varepsilon }{2}$
the situation is inverse and the system chooses the modulation along the
axis ($x$ or $y$) making the largest angle\ with field direction. The
deviation of the modulation direction from the principal axis $x,y$ is small
and for $\beta >\frac{\varepsilon }{2}$ (and $\left\vert \theta \right\vert
<\pi /4$) the equilibrium angle $\varphi $ is 
\begin{equation}
\varphi \approx \frac{\beta +\varepsilon }{\varepsilon q^{2}}(2eH)^{2}\frac{%
d^{2}}{12}\sin 2\theta \approx \beta \frac{\beta +\varepsilon }{\varepsilon
\alpha }(2eH)^{2}\frac{d^{2}}{6}\sin 2\theta \ll 1.  \label{deviation angle}
\end{equation}

The diamagnetic moment of the FFLO state is strongly angular dependent. For $%
\beta >\frac{\varepsilon }{2}$ 
\begin{equation}
M\sim -\left\vert \Psi _{q}\right\vert ^{2}\frac{\alpha }{\beta }Hd^{2}\left[
\beta (1-\left\vert \cos 2\theta \right\vert )+\frac{\varepsilon }{2}%
(1+\left\vert \cos 2\theta \right\vert )\right] ,  \label{Moment}
\end{equation}

Note that in the usual uniform superconducting phase there is no linear
angular dependent contribution to the magnetic moment.

The angular dependence of the critical temperature (critical field) for the
FFLO state is 
\begin{equation}
a(h,T)=\frac{\alpha ^{2}}{2\beta }-\frac{\alpha }{\beta }(2eH)^{2}\frac{d^{2}%
}{12}\left[ \beta +\frac{\varepsilon }{4}-\frac{1}{2}\left( \beta -\frac{%
\varepsilon }{2}\right) \left\vert \cos 2\theta \right\vert \right] ,
\label{Hc2FFLO-1}
\end{equation}%
This angular dependence is presented in Fig. 1. \bigskip In the uniform
phase the angular dependence appears only due to the anisotropy term and it
is proportional to $(2eH)^{4}$ : 
\begin{equation}
a(h,T)=\alpha (2eH)^{2}\frac{d^{2}}{12}-\beta (2eH)^{4}\frac{d^{4}}{80}-%
\frac{\varepsilon }{2}(2eH)^{4}\frac{d^{4}}{80}\frac{1}{2}(1-\cos 4\theta ).
\label{Hc2uniform}
\end{equation}

Comparing the angular dependence of the critical field in FFLO phase (Eq.~%
\ref{Hc2FFLO-1} with that in the uniform state (Eq.~\ref{Hc2uniform}), we
see that the latter is much weaker ($\sim H^{4}$) and has a different form -
see Fig.\ 1.

\begin{figure}[t]
\includegraphics[width=90mm]{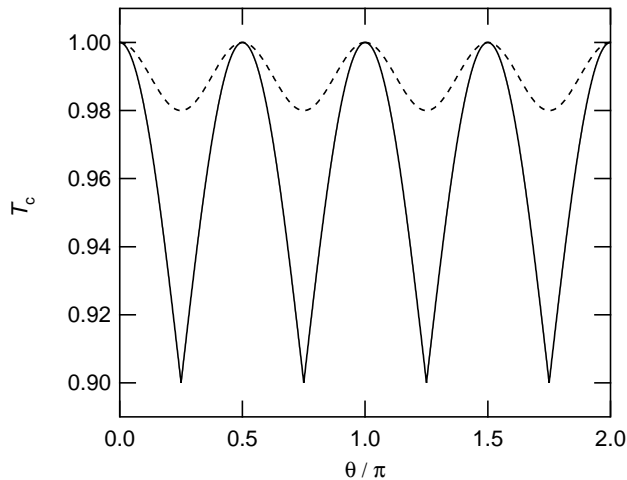}%
\caption{Schematic presentation of the angular dependence of the in-plane
critical field (or critical temperature) in FFLO state (solid line) and in
the usual superconducting state (dotted line).
}
\end{figure}

Therefore the experimental studies of the in plane anisotropy of the
critical field above and below $T^{\ast }$ may provide a conclusive test of
the FFLO state formation. As it has been already noted, the critical field
in CeCoIn$_{5\text{ }}$ is mainly determined by the paramagnetic limit $H$ $%
\sim H_{p}$ (the Maki parameter is large $\alpha _{M}\approx 5$).

The orbital effect sufficiently far away from the tricritical point provides
the relative contribution to critical field of the order of $\sim \left( 
\frac{Hd\xi _{0}}{\Phi _{0}}\right) ^{2}$, where $\xi _{0}$ is the
superconducting correlation length in plane. We may rewrite this as $\left( 
\frac{Hd\xi _{0}}{\Phi _{0}}\right) ^{2}\sim \left( \frac{H_{p}}{H_{orb}}%
\right) ^{2}\left( \frac{d}{\xi _{0}}\right) ^{2}\sim \frac{1}{\left( \alpha
_{M}\right) ^{2}}\left( \frac{d}{\xi _{0}}\right) ^{2}$, with $H_{orb}\sim 
\frac{\Phi _{0}}{\xi _{0}^{2}}.$ Near the tricritical point the usual
orbital effect weakens as $\frac{\alpha }{\alpha _{0}}\left( \frac{H_{p}}{%
H_{orb}}\right) ^{2}\left( \frac{d}{\xi _{0}}\right) ^{2}\sim \frac{\alpha }{%
\alpha _{0}}\frac{1}{\left( \alpha _{M}\right) ^{2}}\left( \frac{d}{\xi _{0}}%
\right) ^{2}$, here $\alpha _{0}$ is the gradient term coefficient near $%
T_{c0}$ (i.e. far away from the FFLO transition). The higher derivatives
terms also contribute to the in-plane anisotropy of the critical field
through the so called non-local corrections to the GL theory \cite{Kogan}.
Their contribution is of the order of $\left( \frac{H_{p}}{H_{orb}}\right)
^{4}\left( \frac{d}{\xi _{0}}\right) ^{4}\sim \frac{1}{\left( \alpha
_{M}\right) ^{4}}\left( \frac{d}{\xi _{0}}\right) ^{4}.$ Therefore the
condition of the domination of the special FFLO behavior being $\frac{\alpha 
}{\alpha _{0}}>\frac{1}{\left( \alpha _{M}\right) ^{2}}\left( \frac{d}{\xi
_{0}}\right) ^{2}.$ Then even for $d\lesssim \xi _{0}$ the characteristic
FFLO regime would be observed everywhere except a tiny vicinity of the
tricritical point.

\qquad In this section we supposed that the FFLO transition is a
second order transition. In the case of the first order transition (like in
CeCoIn$_{5}$\ ) the performed calculations give a field (or
temperature) of the over cooling of the normal phase. The actual field of
the first order transition will be somewhat higher. However in the case of a
weak first order transition the corresponding field (critical temperature)
is obtained by the simple shift of $a(h,T)$ - see next
section. Therefore in this case also we may expect a peculiar angular
dependence (fig. 1) of the critical field.

\section{Current in the FFLO state. Anisotropy of the critical current.}

As it has been discussed in the previous section the orbital effect of the
parallel magnetic field is small but permits to orient the wave-vector of
the FFLO modulation in combination with crystalline/pairing anisotropy and
then the critical current of the film will be anisotropic. Near the
transition into the FFLO state the minimum energy is achieved for the one
dimensional cos-like modulation of the order parameter \cite%
{LarkinOvchin,BuzKach,ShimaharaRainer}. Assuming $\beta >\frac{\varepsilon }{%
2}$ and the magnetic field oriented along the $x$-axis we have 
\begin{equation}
\Psi (\mathbf{r})=f\cos (q_{0}x).  \label{Psi}
\end{equation}%
Apart from the choice of the direction of the FFLO modulation, the orbital
effect of the parallel magnetic field leads only to the small
renormalization of the coefficients of the (\ref{MGL}). Therefore the FFLO
in the thin film opens the possibility to study the critical current and its
anisotropy.

Naturally the ground state $\Psi (\mathbf{r})=f\cos (q_{0}x)$ has no current
and to describe the current carrying states we choose the order parameter in
the form 
\begin{equation}
\Psi (\mathbf{r})=f\cos (q_{0}x)\exp (i\varphi (\mathbf{r})).
\label{Psi gen}
\end{equation}

To calculate the in-plane current we need to introduce the parallel
components of the vector-potential $\mathbf{A}_{\parallel }=\left(
A_{x},A_{y}\right) $ and the part of the functional describing the
interaction with $A_{\parallel }$ being%
\begin{eqnarray}
2\delta F_{\mathbf{A}} &=&a(H,T)f^{2}-\alpha f^{2}\left( \left( \mathbf{%
\nabla }\varphi -2e\mathbf{A}\right) ^{2}+q_{0}^{2}\right) +  \label{F(A)} \\
&&+\beta \left( q_{0}^{4}+6q_{0}^{2}\left( \frac{\partial \varphi }{\partial
x}-2eA_{x}\right) ^{2}+2q_{0}^{2}\left( \frac{\partial \varphi }{\partial y}%
-2eA_{y}\right) ^{2}\right) f^{2}+2\varepsilon q_{0}^{2}\left( \frac{%
\partial \varphi }{\partial y}-2eA_{y}\right) ^{2}f^{2}.  \notag
\end{eqnarray}

In this expression we have retained only the leading gradient terms assuming 
$\left\vert \overrightarrow{\bigtriangledown }\varphi \right\vert <<q_{0%
\text{ }}$and performed the averaging over the FFLO modulation. Note that
for the current calculation we need to use its most general definition $%
\mathbf{j=-}\frac{\delta F}{\delta \mathbf{A}}$ and not the formula from the
Ginzburg-Landau theory. This is a consequence of the fact that the
electrodynamics of the MGL theory is in fact very different from the
standard GL one \cite{Houzet 2001}. Using the relation $q_{0}^{2}=\frac{%
\alpha }{2\beta }$ we obtain the following expressions for the current 
\begin{equation}
j_{x}=4e\alpha f^{2}\left( \frac{\partial \varphi }{\partial x}\right) ,%
\text{ }j_{y}=2e\alpha \frac{\varepsilon }{\beta }f^{2}\left( \frac{\partial
\varphi }{\partial y}\right) .  \label{Current}
\end{equation}

For $\varphi =kx$ the state with the uniform current along the $x$ axis is
realized. It is interesting to note that the perpendicular current \ along
the $y$ axis (with $\varphi =ky$ ) is proportional to the anisotropy
parameter $\varepsilon $ and vanishes in the idealized isotropic model.

Now we calculate the critical current in the FFLO state following the
approach similar to the standard GL theory (see, for example, ref.~%
\onlinecite{De Gennes}) but taking into account the first order character of
the FFLO transition. In the absence of the current the order parameter has
the form (\ref{Psi}) and putting it into (\ref{MGL}) we obtain the averaged
free energy density 
\begin{equation*}
F_{G}=\left( a-a_{2}\right) (H,T)f^{2}-\frac{b}{2}f^{4}+\frac{\lambda }{3}%
f^{6},
\end{equation*}%
where $a=a_{2}=\frac{\alpha ^{2}}{4\beta }$ corresponds to the second order
transition into FFLO state. The first order transition occurs at higher
temperature/magnetic field and its \textquotedblleft temperature" $%
a_{1}=a_{2}+\frac{3b^{2}}{16\lambda }$ and amplitude of the order parameter $%
f_{0}^{2}=\frac{3b^{{}}}{4\lambda }$ may be easily found from the conditions 
\begin{eqnarray}
F_{G}\left( a_{1},f_{0}\right) &=&\left( a_{1}-a_{2}\right) f_{0}^{2}-\frac{b%
}{2}f_{0}^{4}+\frac{\lambda }{3}f_{0}^{6}=0,  \label{equations for Psi0} \\
\frac{\partial F_{G}\left( a_{1},f_{0}\right) }{\partial f_{0}^{2}}
&=&\left( a_{1}-a_{2}\right) -bf_{0}^{2}+\lambda f_{0}^{4}=0.
\label{equations for Psit}
\end{eqnarray}%
Note that the \textquotedblleft critical temperature" \ $a_{1}$%
\ of the fist order transition is simply obtained from the \textquotedblleft
critical temperature" of the second order transition $a_{2}$\ by the
shift on $\frac{3b^{2}}{16\lambda} .$\ 

It is convenient to introduce the normalized \textquotedblleft temperature"
and order parameter: $t=\left( a-a_{2}\right) /\left( a_{1}-a_{2}\right) ,$ $%
\widetilde{f}=f/f_{0}$. Without the current the \textquotedblleft
temperature" dependence of the order parameter is given by $\widetilde{f}%
_{t}^{2}=\left( 2+\sqrt{4-3t}\right) /3.$ In a presence of the current along
the $x$ axis, taking into account the equations (\ref{F(A)}), (\ref{Current}%
), and (\ref{equations for Psit}) we have the following relationship between
the current and the amplitude of the order parameter.%
\begin{equation}
j_{x}^{2}=\frac{3\alpha e^{2}b^{2}}{\lambda }\widetilde{f}%
_{t}^{8}f_{0}^{4}z^{4}(1-z^{2})(7-z^{2}),  \label{jx}
\end{equation}%
where $z=\widetilde{f}/\widetilde{f}_{t}$. The condition of the
applicability of our approach $\left\vert \overrightarrow{\bigtriangledown }%
\varphi \right\vert <<q_{0\text{ }}$reads $b^{2}<<\alpha \lambda /\sqrt{%
\beta }$, i.e. the FFLO transition must be weakly first order. The plot $%
j_{x}(z)$ (see Fig. 2) have a maximum at $z^{2}=\left( 3\sqrt{2}-\sqrt{11}%
\right) /\sqrt{2}\approx 0.65$ which gives us the critical current along the 
$x$ axis 
\begin{equation}
j_{xcrit}^{2}\approx \frac{2.85\alpha e^{2}b^{2}}{\lambda }\widetilde{f}%
_{t}^{8}f_{0}^{4}.  \label{jxcrit}
\end{equation}

\begin{figure}[t]
\includegraphics[width=90mm]{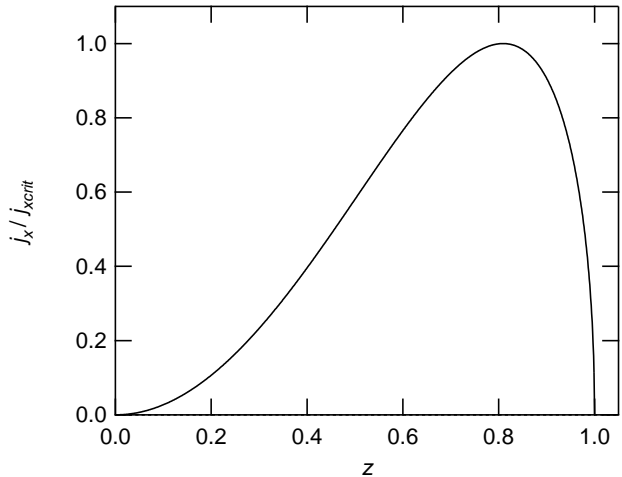}%
\caption{
Superconducting current versus the normalized amplitude of the
order parameter $z=\widetilde{f}/\widetilde{f}_{t}$. Maximum current
corresponds to the critical current of the FFLO state. For both orientations
the maximum of the current is achieved at the same value $z\approx 0.81$.
}
\end{figure}

Analogously we find that the critical current in the direction perpendicular
to the FFLO modulation is 
\begin{equation}
j_{ycrit}^{2}\approx \frac{1}{4}\left( \frac{\varepsilon }{\beta }\right)
^{2}\frac{2.85\alpha e^{2}b^{2}}{\lambda }\widetilde{f}_{t}^{8}f_{0}^{4}.
\label{jycrit}
\end{equation}

The coefficient 1/4 is coming from the averaging on the order parameter
modulation along the $x$ axis. Finally the anisotropy of the critical
current in the FFLO phase is very pronounced 
\begin{equation}
\frac{j_{xcrit}}{j_{ycrit}}=\frac{2\beta }{\varepsilon }.  \notag
\end{equation}%
and in the isotropic model the critical current along the $y$ is even
vanishing. In real system the ratio $\frac{2\beta }{\varepsilon }$ is
expected to be of the order of one and the measurements of the critical
current can permit to determine directly this parameter. As the critical
current in the uniform state is isotropic far away from the trictitical
point, then the experimental observation of the anisotropy of the critical
current may serve as a clear indication of the FFLO phase formation.


\section{Conclusions}

To summarize, we have investigated the properties of the FFLO phase in a
thin film at parallel magnetic field. The orbital effect (even though it is
small) leads to the orientation of the FFLO modulation through the whole
film providing the monodomain FFLO state. Moreover in the FFLO state a
peculiar angular dependence of the in-plane upper critical field must be
observed. This conclusion is quite general and holds for both first and
second order FFLO transitions. We predict also an important anisotropy of
the in-plane critical current in the FFLO phase depending on the current
direction with respect to the FFLO modulation. Such characteristic
anisotropies of the critical field and critical current may be considered as
a smoking gun of the FFLO phase formation. Note that in \cite{Konsch} it has
been demonstrated that the superconducting fluctuational regime changes
drastically near the FFLO TCP. However in the case of the first order FFLO
transition the fluctuational regime could be inaccessible on experiment. Our
analysis was based on the very general MGL functional approach which is
valid for both $s$-wave and $d$-wave superconductors. This approach is fully
justified near the tricritical point and in the case of the weakly first
order phase transition. Nevertheless qualitatively the obtained results
could be extrapolated to the whole region of the FFLO phase existence and
may be relevant for the CeCoIn$_{5}$ thin film experiments. In conclusion we
stress that the predicted anisotropy of the of the in plane critical field
and critical current  must be also observed in  $s$-wave tetragonal FFLO
superconductor.

\section*{Acnkowledgments}

We thank J. Cayssol and M. Houzet for helpful comments. A. B. acknowledges
the support of the JSPS Short-Term Fellowship Program for Research in Japan.





\begin{thebibliography}{99}
\bibitem{Matsuda} Matsuda Y. and Shimahara H. \textit{J. Phys. Soc. Jpn.} 
\textbf{76} (2007) 051005.

\bibitem{LarkinOvchin} Larkin, A. I., and Ovchinnikov Yu. N. \textit{Sov.
Phys. JETP}\ \textbf{20} (1965) 762.

\bibitem{FuldeFerrel} Fulde, P., and Ferrell R. A. \textit{Phys. Rev.} 
\textbf{135}\textit{\ }(1964) A550.\ 

\bibitem{Casalbuoni} Casalbuoni, R., and Nardulli G. \textit{Rev. Mod. Phys. 
} \textbf{76 }(2004) 263.

\bibitem{Uji} Uji S. et al. \textit{Phys. Rev. Lett. }\textbf{97} (2006)
157001.

\bibitem{GunterGrun} Gruenberg L. W. and Gunther L. \textit{Phys. Rev. Lett.}
\textbf{16} (1966) 996.

\bibitem{Bul1973} Bulaevskii L. N. \textit{Sov. Phys. JETP}\ \textbf{38}
(1974) 634.

\bibitem{BuzBrison} Buzdin, A. I. and Brison J.-P. \textit{Phys. Lett. A} \ 
\textbf{218} (1996) 359.

\bibitem{BuzBrison2D} Buzdin, A. I. and Brison J.-P. \textit{Europhys. Lett.}
\ \textbf{35} (1996) 707.

\bibitem{ShimaharaRainer} Shimahara H. and Rainer D. \textit{J. Phys. Soc.
Jpn.} \textbf{66} (1997) 3591.

\bibitem{BuzKach} Buzdin, A. I. and Kachkachi H. \textit{Phys. Lett. A} \ 
\textbf{225} (1997) 341.

\bibitem{Mineev} Mineev V. P. and Samokhin K. V. \textit{Introduction to
unconventional superconductivity of Metals and Alloys} (1999) Gordon and
Breach.

\bibitem{Kogan} Kogan V. G. et al. \textit{Phys. Rev. B }\textbf{62} (2000)
9077.

\bibitem{Izawa} Izawa, K. et al. \textit{Phys. Rev. Lett.} \textbf{87}
(2001) 057002.

\bibitem{Bianchi} Bianchi, A. et al. \textit{Phys. Rev. Lett.} \textbf{91}
(2003) 257001.

\bibitem{Bul-review} Bulaevskii L. N. et al. \textit{Advances in Physics }%
\textbf{34 (}1985) 175.

\bibitem{Samokhin} Samokhin K. V. \textit{Physica C} \textbf{274} (1997) 156.

\bibitem{Yang} Yang K. and Sondhi S. L. \textit{Phys. Rev. B }\textbf{57}
(1998) 8566.

\bibitem{Shimahara1} Shimahara H. \textit{J. Phys. Soc. Jpn.} \textbf{66}
(1997) 541.

\bibitem{Shimahara2} Shimahara H. \textit{J. Phys. Soc. Jpn.} \textbf{68}
(1999) 3069.

\bibitem{Brison} Brison J. P. et al. \textit{Physica C} \textbf{250} (1995)
128.

\bibitem{Houzet} Houzet, M. and Mineev V. P. \textit{Phys. Rev. B }\textbf{74%
} (2006) 144522.

\bibitem{Houzet 2001} Houzet, M. and Buzdin A. \textit{Phys. Rev. B }\textbf{%
63} (2001) 184521.

\bibitem{De Gennes} De Gennes, P. G. \textit{Superconductivity of Metals and
Alloys} (1966) New York : Benjamin.

\bibitem{Konsch} Konschelle, F. et al. \textit{Europhys. Lett.} \ \textbf{79}
(2007) 67001.
\end{thebibliography}
\end{document}